\documentclass[11pt,onecolumn,draftclsnofoot]{IEEEtran}

\usepackage{amsmath,amssymb} \interdisplaylinepenalty=2500
\usepackage{cite}

\newtheorem{theorem}{Theorem}
\newtheorem{proposition}[theorem]{Proposition}
\newtheorem{corollary}[theorem]{Corollary}
\newtheorem{lemma}[theorem]{Lemma}

\newenvironment{remark}{\textit{Remark: }}{}

\renewcommand{\dim}{\operatorname{\sf dim}\hspace{0.1em}}
\DeclareMathOperator{\rank}{\sf rank\hspace{0.1em}}
\newcommand{\RRE}[1]{{\sf RRE}\left(#1\right)}

\newcommand{\Fq}{\mathbb{F}_q}
\newcommand{\mat}[1]{\begin{bmatrix} #1 \end{bmatrix}}
\newcommand{\gc}[2]{\genfrac{[}{]}{0pt}{}{#1}{#2}}

\newcommand{\calC}{\mathcal{C}}

\newcommand{\calT}{\mathcal{T}}

\newcommand{\calX}{\mathcal{X}}
\newcommand{\calY}{\mathcal{Y}}

\newcommand{\AUXVAR}{Q}
\newcommand{\DATAMATRIX}{U}

\title{Communication over Finite-Field\\ Matrix Channels}

\author{Danilo Silva, Frank R. Kschischang, and Ralf K\"{o}tter
\thanks{This work was supported by CAPES Foundation, Brazil, and by the Natural Sciences and Engineering Research Council of Canada. The material in this paper was presented in part at the 24th Biennial Symposium on Communications, Kingston, Canada, June 2008.}
\thanks{D. Silva and F. R. Kschischang are with The Edward S. Rogers Sr. Department of Electrical and Computer Engineering, University of Toronto, Toronto, ON M5S 3G4, Canada (e-mail: danilo@comm.utoronto.ca, frank@comm.utoronto.ca).}
\thanks{R. K\"{o}tter is deceased.
}
}

\begin{document}
\maketitle
\thispagestyle{empty}

\begin{abstract}

This paper is motivated by the problem of error control in network coding when errors are introduced in a random fashion (rather than chosen by an adversary). An additive-multiplicative matrix channel is considered as a model for random network coding. The model assumes that $n$ packets of length $m$ are transmitted over the network, and up to $t$ erroneous packets are randomly chosen and injected into the network. Upper and lower bounds on capacity are obtained for any channel parameters, and asymptotic expressions are provided in the limit of large field or matrix size. A simple coding scheme is presented that achieves capacity in both limiting cases. The scheme has decoding complexity $O(n^2 m)$ and a probability of error that decreases exponentially both in the packet length and in the field size in bits. Extensions of these results for coherent network coding are also presented.

\end{abstract}

\begin{IEEEkeywords}
Error correction, error trapping, matrix channels, network coding, one-shot codes, probabilistic error model.
\end{IEEEkeywords}

\section{Introduction}
\label{sec:introduction}

Linear network coding \cite{Ahlswede++2000,Li++2003,Koetter.Medard2003} is a promising new approach to information dissemination over networks. The fact that packets may be linearly combined at intermediate nodes affords, in many useful scenarios, higher rates than conventional routing approaches. If the linear combinations are chosen in a random, distributed fashion, then random linear network coding \cite{Ho++2006} not only maintains most of the benefits of linear network coding, but also affords a remarkable simplicity of design that is practically very appealing.

However, linear network coding has the intrinsic drawback of being extremely sensitive to error propagation. Due to packet mixing, a single corrupt packet has the potential to contaminate all packets received by a destination node. The problem is better understood by looking at a matrix model for (single-source) linear network coding, given by
\begin{equation}\label{eq:basic-channel-model}
  Y = AX + DZ.
\end{equation}
All matrices are over a finite field. Here, $X$ is an $n \times m$ matrix whose rows are packets transmitted by the source node, $Y$ is an $N \times m$ matrix whose rows are the packets received by a (specific) destination node, and $Z$ is a $t \times m$ matrix whose rows are the additive error packets injected at some network links. The matrices $A$ and $D$ are transfer matrices that describe the linear transformations incurred by packets on route to the destination. Such linear transformations are responsible for the (unconventional) phenomenon of error propagation.

There has been an increasing amount of research on error control for network coding, with results naturally depending on the specific channel model used, i.e., the joint statistics of $A$, $D$ and $Z$ given $X$. Under a worst-case (or adversarial) error model, the work in \cite{Silva.Kschischang2008:Metrics,Silva.Kschischang2008:Allerton} (together with \cite{Cai.Yeung2006,Yeung.Cai2006,Kotter.Kschischang2008,Silva++2008}) has obtained the maximum achievable rate for a wide range of conditions. If $A$ is square ($N=n$) and nonsingular, and $m \geq n$, then the maximum information rate that can be achieved in a single use of the channel is exactly $n-2t$ packets when $A$ is known at the receiver, and approximately $\frac{m-n}{m}(n-2t)$ packets when $A$ is unknown. These approaches are inherently pessimistic and share many similarities with classical coding theory.

Recently, Montanari and Urbanke \cite{Montanari.Urbanke2007} brought the problem to the realm of information theory by considering a probabilistic error model. Their model assumes, as above, that $A$ is invertible and $m \geq n$; in addition, they assume that the matrix $DZ$ is chosen uniformly at random among all $n \times m$ matrices of rank $t$.
For such a model and, under the assumption that the transmitted matrix $X$ must contain an $n \times n$ identity submatrix as a header, they compute the maximal mutual information in the limit of large matrix size---approximately $\frac{m-n-t}{m}(n-t)$ packets per channel use. They also present an iterative coding scheme with decoding complexity $O(n^3m)$ that asymptotically achieves this rate.

The present paper is motivated by \cite{Montanari.Urbanke2007}, and by the challenge of computing or approximating the actual channel capacity (i.e., without any prior assumption on the input distribution) for any channel parameters (i.e., not necessarily in the limit of large matrix size).
Our contributions can be summarized as follows:
\begin{itemize}
  \item Assuming that the matrix $A$ is a constant known to the receiver, we compute the exact channel capacity for any channel parameters. We also present a simple coding scheme that asymptotically achieves capacity in the limit of large field or matrix size.
  \item Assuming that the matrix $A$ is chosen uniformly at random among all nonsingular matrices, we compute upper and lower bounds on the channel capacity for any channel parameters. These bounds are shown to converge asymptotically in the limit of large field or matrix size. We also present a simple coding scheme that asymptotically achieves capacity in both limiting cases. The scheme has decoding complexity $O(n^2 m)$ and a probability of error that decays exponentially fast both in the packet length and in the field size in bits.
  \item We present several extensions of our results for situations where the matrices $A$, $D$ and $Z$ may be chosen according to more general probability distributions.
\end{itemize}

A main assumption that underlies this paper (even the extensions mentioned above) is that the transfer matrix $A$ is always invertible. One might question whether this assumption is realistic for actual network coding systems. For instance, if the field size is small, then random network coding may not produce a nonsingular $A$ with high probability. We believe, however, that removing this assumption complicates the analysis without offering much insight. Under an \emph{end-to-end coding} (or \emph{layered}) approach, there is a clear separation between the network coding protocol---which induces a matrix channel---and the error control techniques applied at the source and destination nodes. In this case, it is reasonable to assume that network coding system will be designed to be \emph{feasible} (i.e., able to deliver $X$ to all destinations) when no errors occur in the network. Indeed, a main premise of linear network coding is that the field size is sufficiently large in order to allow a feasible network code. Thus, the results of this paper may be seen as conditional on the network coding layer being successful in its task.

The remainder of this paper is organized as follows. In Section~\ref{sec:matrix-channels}, we provide general considerations on the type of channels studied in this paper. In Section~\ref{sec:MMC}, we address a special case of (\ref{eq:basic-channel-model}) where $A$ is random and $t=0$, which may be seen as a model for random network coding without errors. In Section~\ref{sec:AMC}, we address a special case of (\ref{eq:basic-channel-model}) where $A$ is the identity matrix. This channel may be seen as a model for network coding with errors when $A$ is known at the receiver, since the receiver can always compute $A^{-1} Y$. The complete channel with a random, unknown $A$ is addressed Section~\ref{sec:AMMC}, where we make crucial use of the results and intuition developed in the previous sections. Section~\ref{sec:extensions} discusses possible extensions of our results, and Section~\ref{sec:conclusion} presents our conclusions.

We will make use of the following notation. Let $\Fq$ be the finite field with $q$ elements. We use $\Fq^{n \times m}$ to denote the set of all $n \times m$ matrices over $\Fq$ and $\calT_{n \times m,t}(\Fq)$ to denote the set of all $n \times m$ matrices of rank $t$ over $\Fq$. We shall write simply $\calT_{n \times m,t} = \calT_{n \times m,t}(\Fq)$ when the field $\Fq$ is clear from the context. We also use the notation $\calT_{n \times m} = \calT_{n \times m,\min\{n,m\}}$ for the set of all full-rank $n \times m$ matrices. The $n \times m$ all-zero matrix and the $n \times n$ identity matrix are denoted by $0_{n \times m}$ and $I_{n \times n}$, respectively, where the subscripts may be omitted when there is no risk of confusion. The reduced row echelon (RRE) form of a matrix $M$ will be denoted by $\RRE{M}$.

\section{Matrix Channels}
\label{sec:matrix-channels}

For clarity and consistency of notation, we recall a few definitions from information theory \cite{Cover.Thomas}.

A discrete channel $(\calX,\calY,p_{Y|X})$ consists of an input alphabet $\calX$, an output alphabet $\calY$, and a conditional probability distribution $p_{Y|X}$ relating the channel input $X \in \calX$ and the channel output $Y \in \calY$. An $(M,\ell)$ code for a channel $(\calX,\calY,p_{Y|X})$ consists of an encoding function $\{1,\ldots,M\} \to \calX^\ell$ and a decoding function $\calY^\ell \to \{1,\ldots,M,f\}$, where $f$ denotes a decoding failure. It is understood that an $(M,\ell)$ code is applied to the $\ell$th extension of the discrete memoryless channel $(\calX,\calY,p_{Y|X})$. A rate $R$ (in bits) is said to be achievable if there exists a sequence of $(\lceil 2^{\ell R} \rceil,\ell)$ codes such that decoding is unsuccessful (either an error or a failure occurs) with probability arbitrarily small as $\ell \to \infty$. The capacity of the channel is the supremum of all achievable rates. It is well-known that the capacity is given by
\begin{equation}\nonumber
  C = \max_{p_X}\, I(X;Y)
\end{equation}
where $p_X$ denotes the input distribution.

Here, we are interested in matrix channels, i.e., channels for which both the input and output variables are matrices. In particular, we are interested in a family of additive matrix channels given by the channel law
\begin{equation}\label{eq:prob-basic-channel-law}
  Y = AX + DZ
\end{equation}
where $X,Y \in \Fq^{n \times m}$, $A \in \Fq^{n \times n}$, $D \in \Fq^{n \times t}$, $Z \in \Fq^{t \times m}$, and $X$, $(A,D)$ and $Z$ are statistically independent. Since the capacity of a matrix channel naturally scales with $nm$, we also define a \emph{normalized capacity}
\begin{equation}\nonumber
  \overline{C} = \frac{1}{nm} C.
\end{equation}

In the following, we assume that statistics of $A$, $D$ and $Z$ are given for all $q,n,m,t$. In this case, we may denote a matrix channel simply by the tuple $(q,n,m,t)$, and we may also indicate this dependency in both $C$ and $\overline{C}$. We now define two limiting forms of a matrix channel (strictly speaking, of a sequence of matrix channels). The first form, which we call the \emph{infinite-field-size channel}, is obtained by taking $q \to \infty$. The capacity of this channel is given by
\begin{equation}\nonumber
  \lim_{q \to \infty} \frac{1}{\log_2 q} C(q,n,m,t)
\end{equation}
represented in $q$-ary units per channel use. The second form, which we call the \emph{infinite-rank channel}, is obtained by setting $t = \tau n$ and $n = \lambda m$, and taking $m \to \infty$. The normalized capacity of this channel is given by
\begin{equation}\nonumber
  \lim_{m \to \infty} \frac{1}{\log_2 q} \overline{C}(q, \lambda m, m, \tau \lambda m)
\end{equation}
represented in $q$-ary units per transmitted $q$-ary symbol. We will hereafter assume that logarithms are taken to the base $q$ and omit the factor $\frac{1}{\log_2 q}$ from the above expressions.

Note that, to achieve the capacity of an infinite-field-size channel (similarly for an infinite-rank channel), one should find a two-dimensional family of codes: namely, a sequence of codes with increasing block length $\ell$ for each $q$, as $q \to \infty$ (or for each $m$, as $m \to \infty$).

We will simplify our task here by considering only codes with block length $\ell=1$, which we call \emph{one-shot codes}. We will show, however, that these codes can achieve the capacity of both the infinite-field-size and the infinite-rank channels, at least for the classes of channels considered here. In other words, one-shot codes are asymptotically optimal as either $q \to \infty$ or $m \to \infty$.

For completeness, we define also two more versions of the channel: the \emph{infinite-packet-length channel}, obtained by fixing $q$, $t$ and $n$, and letting $m \to \infty$, and the \emph{infinite-batch-size channel}, obtained by fixing $q$, $t$ and $m$, and letting $n \to \infty$. These channels are discussed in Section~\ref{ssec:other-infinite-channels}.

It is important to note that a $(q,n,\ell m, t)$ channel is not the same as the $\ell$-extension of a $(q,n,m,t)$ channel. For instance, the $2$-extension of a $(q,n,m,t)$ channel has channel law
\begin{equation}\nonumber
  (Y_1,\, Y_2) = (A_1 X_1 + D_1 Z_1,\, A_2 X_2 + D_2 Z_2)
\end{equation}
where $(X_1, X_2) \in \left(\Fq^{n \times m}\right)^2$, and $(A_1,D_1,Z_1)$ and $(A_2,D_2,Z_2)$ correspond to independent realizations of a $(q,n,m,t)$ channel. This is not the same as the channel law for a $(q,n,2m,t)$ channel,
\begin{equation}\nonumber
  \mat{Y_1 & Y_2} = A_1 \mat{X_1 & X_2} + D_1 \mat{Z_1 & Z_2}
\end{equation}
since $(A_2,D_2)$ may not be equal to $(A_1,D_1)$.

To the best of our knowledge, the $\ell$-extension of a $(q,n,m,t)$ channel has not been considered in previous works, with the exception of \cite{Siavoshani++2008}. For instance, \cite{Jaggi++2008} and \cite{Montanari.Urbanke2007} consider only limiting forms of a $(q,n,m,t)$ channel. Although both models are referred to simply as ``random linear network coding,'' the model implied by the results in \cite{Montanari.Urbanke2007} is in fact an infinite-rank channel, while the model implied by the results in \cite{Jaggi++2008} is an infinite-packet-length-infinite-field-size channel.


We now proceed to investigating special cases of (\ref{eq:prob-basic-channel-law}), by considering specific statistics for $A$, $D$ and $Z$.

\section{The Multiplicative Matrix Channel}
\label{sec:MMC}

We define the \emph{multiplicative matrix channel} (MMC) by the channel law
\begin{equation}\nonumber
  Y = AX
\end{equation}
where $A \in \calT_{n \times n}$ is chosen uniformly at random among all $n \times n$ nonsingular matrices, and independently from $X$.
Note that the MMC is a $(q,n,m,0)$ channel.

\subsection{Capacity and Capacity-Achieving Codes}

In order to find the capacity of this channel, we will first solve a more general problem.

\begin{proposition}\label{prop:MMC-group-capacity}
  Let $\mathcal{G}$ be a finite group that acts on a finite set $\mathcal{S}$. Consider a channel with input variable $X \in \mathcal{S}$ and output variable $Y \in \mathcal{S}$ given by $Y = AX$, where $A \in \mathcal{G}$ is drawn uniformly at random and independently from $X$. The capacity of this channel, in bits per channel use, is given by
\begin{equation}\nonumber
  C = \log_2 |\mathcal{S}/\mathcal{G}|
\end{equation}
where $|\mathcal{S}/\mathcal{G}|$ is the number of equivalence classes of $\mathcal{S}$ under the action of $\mathcal{G}$. Any complete set of representatives of the equivalence classes is a capacity-achieving code.
\end{proposition}
\begin{proof}
For each $x \in \mathcal{S}$, let $\mathcal{G}(x) = \{gx \mid g \in \mathcal{G}\}$ denote the orbit of $x$ under the action of $\mathcal{G}$. Recall that $\mathcal{G}(y) = \mathcal{G}(x)$ for all $y \in \mathcal{G}(x)$ and all $x \in \mathcal{S}$, that is, the orbits form equivalence classes.

For $y \in \mathcal{G}(x)$, let $\mathcal{G}_{x,y} = \{g \in \mathcal{G} \mid gx=y\}$. By a few manipulations, it is easy to show that $|\mathcal{G}_{x,y}| = |\mathcal{G}_{x,y'}|$ for all $y,y' \in \mathcal{G}(x)$. Since $A$ has a uniform distribution, it follows that $P[Y=y \mid X=x] = 1/|\mathcal{G}(x)|$, for all $y \in \mathcal{G}(x)$.

For any $x \in \mathcal{S}$, consider the same channel but with the input alphabet restricted to $\mathcal{G}(x)$. Note that the output alphabet will also be restricted to $\mathcal{G}(x)$. This is a $|\mathcal{G}(x)|$-ary channel with uniform transition probabilities; thus, the capacity of this channel is 0. Now, the overall channel can be considered as a sum (union of alphabets) of all the restricted channels. The capacity of a sum of $M$ channels with capacities $C_i$, $i=1,\ldots,M$, is known to be $\log_2 \sum_{i=1}^M 2^{C_i}$ bits. Thus, the capacity of the overall channel is $\log_2 M$ bits, where $M=|\mathcal{S}/\mathcal{G}|$ is the number of orbits. A capacity-achieving code (with block length 1) may be obtained by simply selecting one representative from each equivalence class.
\end{proof}

Proposition~\ref{prop:MMC-group-capacity} shows that in a channel induced by a group action, where the group elements are selected uniformly at random, the receiver cannot distinguish between transmitted elements that belong to the same equivalence class. Thus, the transmitter can only communicate the choice of a particular equivalence class.

Returning to our original problem, we have $\mathcal{S} = \Fq^{n \times m}$ and $\mathcal{G} = \calT_{n \times n}$ (the general linear group $GL_n(\Fq)$). The equivalence classes of $\mathcal{S}$ under the action of $\mathcal{G}$ are the sets of matrices that share the same row space. Thus, we can identify each equivalence class with a subspace of $\Fq^m$ of dimension at most $n$.
Let the Gaussian coefficient
\begin{equation}\nonumber
  \gc{m}{k}_q = {\prod_{i=0}^{k-1} (q^m - q^i)}/{(q^k - q^i)}
\end{equation}
denote the number of $k$-dimensional subspaces of $\Fq^m$.
We have the following corollary of Proposition~\ref{prop:MMC-group-capacity}.

\begin{corollary}\label{cor:MMC-capacity}
  The capacity of the MMC, in $q$-ary units per channel use, is given by
\begin{equation}\nonumber
 C_{\text{\rm MMC}} = \log_q \sum_{k=0}^n \gc{m}{k}_q.
\end{equation}
A capacity-achieving code $\calC \subseteq \Fq^{n \times m}$ can be obtained by ensuring that each $k$-dimensional subspace of $\Fq^m$, $k \leq n$, is the row space of some unique $X \in \calC$.
\end{corollary}

Note that Corollary~\ref{cor:MMC-capacity} reinforces the idea introduced in \cite{Kotter.Kschischang2008} that, in order to communicate under random network coding, the transmitter should encode information in the choice of a subspace.

We now compute the capacity for the two limiting forms of the channel, as discussed in Section~\ref{sec:matrix-channels}. We have the following result.

\begin{proposition}\label{prop:MMC-capacity-limit}
  Let $\lambda = n/m$ and assume $0 < \lambda \leq 1/2$. Then
\begin{align}
\lim_{q \to \infty} C_{\text{\rm MMC}} &= (m-n)n \label{eq:MMC-capacity-limit-q} \\
\lim_{\substack{m \to \infty \\ n = \lambda m}} \overline{C_{\text{\rm MMC}}} &= 1 - \lambda. \label{eq:MMC-capacity-limit-m}
\end{align}
\end{proposition}
\begin{proof}
First, observe that
\begin{equation}\label{eq:bound-sum-gc}
\gc{m}{n^*}_q < \sum_{k=0}^n \gc{m}{k}_q < (n+1) \gc{m}{n^*}_q
\end{equation}
where $n^* = \min\{n,\,\lfloor m/2 \rfloor\}$. Using the fact that \cite{Kotter.Kschischang2008}
\begin{equation}\label{eq:gauss-coeff-bound}
 q^{(m-k)k} < \gc{m}{k}_q < 4 q^{(m-k)k}
\end{equation}
it follows that
\begin{equation}\label{eq:MMC-capacity-general}
(m-n^*)n^* < C_{\text{\rm MMC}} < (m-n^*)n^* + \log_q 4(n+1).
\end{equation}
The last term on the right vanishes on both limiting cases.
\end{proof}

The case $\lambda \geq 1/2$ can also be readily obtained but is less interesting since, in practice, the packet length $m$ will be much larger than the number of packets $n$.

Note that an expression similar to (\ref{eq:MMC-capacity-general}) has been found in \cite{Siavoshani++2008} under a different assumption on the transfer matrix (namely, that $A$ is uniform on $\Fq^{n \times n})$. It is interesting to note that, also in that case, the same conclusion can be reached about the sufficiency of transmitting subspaces \cite{Siavoshani++2008}.

An intuitive way to interpret (\ref{eq:MMC-capacity-limit-q}) is the following: out of the $nm$ symbols obtained by the receiver, $n^2$ of these symbols are used to describe $A$, while the remaining ones are used to communicate $X$.

It is interesting to note that (\ref{eq:MMC-capacity-limit-q}) precisely matches (\ref{eq:MMC-capacity-limit-m}) after normalizing by the total number of transmitted symbols, $nm$.

Both limiting capacity expressions (\ref{eq:MMC-capacity-limit-q}) and (\ref{eq:MMC-capacity-limit-m}) can be achieved using a simple coding scheme where an $n \times (m-n)$ data matrix $\DATAMATRIX$ is concatenated on the left with an $n \times n$ identity matrix $I$, yielding a transmitted matrix $X = \mat{I & \DATAMATRIX}$. The first $n$ symbols of each transmitted packet may be interpreted as pilot symbols used to perform ``channel sounding''. Note that this is simply the standard way of using random network coding \cite{Chou++2003}.

\section{The Additive Matrix Channel}
\label{sec:AMC}

We define the \emph{additive matrix channel} (AMC) according to
\begin{equation}\nonumber
  Y = X + W
\end{equation}
where $W \in \calT_{n \times m,t}$ is chosen uniformly at random among all $n \times m$ matrices of rank $t$, independently from $X$.
Note that the AMC is a $(q,n,m,t)$ channel with $D \in \calT_{n \times t}$ and $Z \in \calT_{t \times m}$ uniformly distributed, and $A = I$.

\subsection{Capacity}

The capacity of the AMC is computed in the next proposition.

\begin{proposition}\label{prop:AMC-capacity}
The capacity of the AMC is given by
\begin{equation}\nonumber
  C_{\text{\rm AMC}} = nm - \log_q |\calT_{n\times m,t}|.
\end{equation}
For $\lambda = n/m$ and $\tau = t/n$, we have the limiting expressions
\begin{align}
\lim_{q \to \infty} C_{\text{\rm AMC}} &= (m - t)(n-t) \label{eq:AMC-capacity-limit-q} \\
\lim_{\substack{m \to \infty \\ n = \lambda m \\ t = \tau n}} \overline{C_{\text{\rm AMC}}} &= (1 - \lambda \tau)(1-\tau). \label{eq:AMC-capacity-limit-m}
\end{align}
\end{proposition}
\begin{proof}
To compute the capacity, we expand the mutual information
\begin{equation}\nonumber
  I(X;Y) = H(Y) - H(Y|X) = H(Y) - H(W)
\end{equation}
where the last equality holds because $X$ and $W$ are independent. Note that $H(Y) \leq nm$, and the maximum is achieved when $Y$ is uniform. Since $H(W)$ does not depend on the input distribution, we can maximize $H(Y)$ by choosing, e.g., a uniform $p_X$.

The entropy of $W$ is given by $H(W) = \log_q |\calT_{n\times m,t}|$. The number of $n \times m$ matrices of rank $t$ is given by \cite[p. 455]{Lidl.Niederreiter}
\begin{align}
|\calT_{n \times m,t}|
&= \frac{|\calT_{n \times t}||\calT_{t \times m}|}{|\calT_{t \times t}|} = |\calT_{n \times t}| \gc{m}{t}_q \label{eq:number-matrices-rank} \\
&= q^{(n+m-t)t}\prod_{i=0}^{t-1}\frac{(1 - q^{i-n})(1 - q^{i-m})}{(1 - q^{i-t})}. \label{eq:number-matrices-rank-2}
\end{align}

Thus,
\begin{align}
C_{\text{\rm AMC}}
&= nm - \log_q |\calT_{n\times m,t}| \nonumber \\
&= (m - t)(n-t) + \log_q \prod_{i=0}^{t-1}\frac{(1 - q^{i-t})}{(1 - q^{i-n})(1 - q^{i-m})}  \nonumber
\end{align}
The limiting expressions (\ref{eq:AMC-capacity-limit-q}) and (\ref{eq:AMC-capacity-limit-m}) follow immediately from the equation above.
\end{proof}

\begin{remark}
  The expression (\ref{eq:AMC-capacity-limit-m}), which gives the capacity of the infinite-rank AMC, has been previously obtained in \cite{Montanari.Urbanke2007} for a channel that is equivalent to the AMC. Our proof is a simple extension of the proof in \cite{Montanari.Urbanke2007}.
\end{remark}

As can be seen from (\ref{eq:number-matrices-rank-2}), an $n \times m$ matrix of rank $t$ can be specified with approximately $(n+m-t)t$ symbols. Thus, the capacity (\ref{eq:AMC-capacity-limit-q}) can be interpreted as the number of symbols conveyed by $Y$ minus the number of symbols needed to describe $W$.

Note that, as in Section~\ref{sec:MMC}, the normalized capacities of the infinite-field-size AMC and the infinite-rank AMC are the same. An intuitive explanation might be the fact that, for the two channels, both the number of bits per row and the number of bits per column tend to infinity. In contrast, the normalized capacity is different when only one of these quantities grows while the other is fixed. This is the case of the infinite-packet-length AMC and the infinite-batch-size AMC, which are studied in Section~\ref{ssec:other-infinite-channels}.

\subsection{A Coding Scheme}

We now present an efficient coding scheme that achieves (\ref{eq:AMC-capacity-limit-q}) and (\ref{eq:AMC-capacity-limit-m}). The scheme is based on an ``error trapping'' strategy.

Let $\DATAMATRIX \in \Fq^{(n-v) \times (m-v)}$ be a data matrix, where $v \geq t$. A codeword $X$ is formed by adding all-zero rows and columns to $\DATAMATRIX$ so that
\begin{equation}\nonumber
  X = \mat{0_{v \times v} & 0_{v \times (m-v)} \\ 0_{(n-v) \times v} & \DATAMATRIX}.
\end{equation}
These all-zero rows and columns may be interpreted as the ``error traps.''
Clearly, the rate of this scheme is $R = (n-v)(m-v)$.

Since the noise matrix $W$ has rank $t$, we can write it as
\begin{equation}\nonumber
  W = BZ = \mat{B_1 \\ B_2} \mat{Z_1 & Z_2}
\end{equation}
where $B_1 \in \Fq^{v \times t}$, $B_2 \in \Fq^{(n-v) \times t}$, $Z_1 \in \Fq^{t \times v}$ and $Z_2 \in \Fq^{t \times (m-v)}$.
The received matrix $Y$ is then given by
\begin{equation}\nonumber
  Y = X + W = \mat{B_1 Z_1 & B_1 Z_2 \\ B_2 Z_1 & \DATAMATRIX + B_2 Z_2}.
\end{equation}

We define an error trapping failure to be the event that $\rank B_1 Z_1 < t$. Intuitively, this corresponds to the situation where either the row space or the column space of the error matrix has not been ``trapped''.

For now, assume that the error trapping is successful, i.e., $\rank B_1 = \rank Z_1 = t$. Consider the submatrix corresponding to the first $v$ columns of $Y$. Since $\rank B_1 Z_1 = t$, the rows of $B_2 Z_1$ are completely spanned by the rows of $B_1 Z_1$. Thus, there exists some matrix $\bar{T}$ such that $B_2 Z_1 = \bar{T} B_1 Z_1$. But $(B_2 - \bar{T}B_1)Z_1 = 0$ implies that $B_2 - \bar{T}B_1 = 0$, since $Z_1$ has full row rank. It follows that
\begin{equation}\nonumber
  T \mat{B_1 \\ B_2} = \mat{B_1 \\ 0},\, \text{ where }\, T = \mat{I & 0 \\ \bar{T} & I}.
\end{equation}
Note also that $TX = X$. Thus,
\begin{equation}\nonumber
  TY = TX + TW = \mat{B_1 Z_1 & B_1 Z_2 \\ 0 & \DATAMATRIX}
\end{equation}
from which the data matrix $\DATAMATRIX$ can be readily obtained.

The complexity of the scheme is computed as follows. In order to obtain $\bar{T}$, it suffices to perform Gaussian elimination on the left $n \times v$ submatrix of $Y$, for a cost of $O(nv^2)$ operations. The data matrix can be extracted by multiplying $\bar{T}$ with the top right $v \times (n-v)$ submatrix of $Y$, which can be accomplished in $O((n-v)v(m-v))$ operations. Thus, the overall complexity of the scheme is $O(nmv)$ operations in $\Fq$.

Note that $B_1 Z_1$ is available at the receiver as the top-left submatrix of $Y$. Moreover, the rank of $B_1 Z_1$ is already computed during the Gaussian elimination step of the decoding. Thus, the event that the error trapping fails can be readily detected at the receiver, which can then declare a decoding failure. It follows that the error probability of the scheme is zero.

Let us now compute the probability of decoding failure. Consider, for instance, $P_1 = P[\rank Z_1 = t]$, where $Z = \mat{Z_1 & Z_2}$ is a full-rank matrix chosen uniformly at random. An equivalent way of generating $Z$ is to first generate the entries of a matrix $M \in \Fq^{t \times m}$ uniformly at random, and then discard $M$ if it is not full-rank. Thus, we want to compute $P_1 = P[\rank M_1 = t \mid \rank M = t]$, where $M_1$ corresponds to the first $v$ columns of $M$. This probability is
\begin{align}
  P_1
&= \frac{P[\rank M_1 = t]}{P[\rank M = t]} = \frac{q^{mt}\prod_{i=0}^{t-1} (q^v - q^i)}{q^{vt}\prod_{i=0}^{t-1} (q^m - q^i)} \nonumber \\
&> \prod_{i=0}^{t-1} (1 - q^{i-v}) \geq (1 - q^{t-1-v})^t \geq 1 - \frac{t}{q^{1+v-t}}. \nonumber
\end{align}
The same analysis holds for $P_2 = P[\rank B_1 = t]$. By the union bound, it follows that the probability of failure satisfies
\begin{equation}\label{eq:AMC-failure-prob}
  P_f < \frac{2t}{q^{1 + v-t}}.
\end{equation}

\begin{proposition}\label{prop:AMC-coding-achievability}
  The coding scheme described above can achieve both capacity expressions (\ref{eq:AMC-capacity-limit-q}) and (\ref{eq:AMC-capacity-limit-m}).
\end{proposition}
\begin{proof}
From (\ref{eq:AMC-failure-prob}), we see that achieving either of the limiting capacities amounts to setting a suitable $v$. To achieve (\ref{eq:AMC-capacity-limit-q}), we set $v=t$ and let $q$ grow. The resulting code will have the correct rate, namely, $R = (n-t)(m-t)$ in $q$-ary units, while the probability of failure will decrease exponentially with the field size in bits.

Alternatively, to achieve (\ref{eq:AMC-capacity-limit-m}), we can choose some small $\epsilon > 0$ and set $v = (\tau + \epsilon) n$, where both $\tau = t/n$ and $\lambda = n/m$ are assumed fixed. By letting $m$ grow, we obtain a probability of failure that decreases exponentially with $m$. The (normalized) gap to capacity of the resulting code will be
\begin{align}
\bar{g}
&\triangleq \lim_{m \to \infty}\, \overline{C_{\text{\rm AMC}}}-R/(nm) \nonumber \\
&= (1-\lambda\tau)(1-\tau) - (1 - \lambda(\tau + \epsilon))(1 - (\tau + \epsilon)) \nonumber \\
&= \lambda\epsilon(1-(\tau+\epsilon)) + \epsilon(1-\lambda\tau) \nonumber \\
&< \lambda\epsilon + \epsilon = (1+\lambda)\epsilon \nonumber
\end{align}
which can be made as small as we wish.
\end{proof}

\section{The Additive-Multiplicative Matrix Channel}
\label{sec:AMMC}

Consider a $(q,n,m,t)$ channel with $A \in \calT_{n \times n}$, $D \in \calT_{n \times t}$ and $Z \in \calT_{t \times m}$ uniformly distributed and independent from other variables.
Since $A$ is invertible, we can rewrite (\ref{eq:prob-basic-channel-law}) as
\begin{equation}\label{eq:AMMC-model}
  Y = AX + DZ = A(X + A^{-1}DZ).
\end{equation}
Now, since $\calT_{n \times n}$ acts transitively on $\calT_{n \times t}$, the channel law (\ref{eq:AMMC-model}) is equivalent to
\begin{equation}\label{eq:AMMC-model-2}
  Y = A(X + W)
\end{equation}
where $A \in \calT_{n \times n}$ and $W \in \calT_{n \times m,t}$ are chosen uniformly at random and independently from any other variables. We call (\ref{eq:AMMC-model-2}) \emph{the additive-multiplicative matrix channel} (AMMC).

\subsection{Capacity}

 One of the main results of this section is the following theorem, which provides an upper bound on the capacity of the AMMC.

\begin{theorem} \label{thm:AMMC-capacity-upper-bound}
For $n \leq m/2$, the capacity of the AMMC is upper bounded by
\begin{equation}\nonumber
  C_{\text{\rm AMMC}} \leq (m-n)(n-t) + \log_q 4(1+n)(1+t).
\end{equation}
\end{theorem}
\begin{proof}
Let $S = X+W$. By expanding $I(X,S;Y)$, and using the fact that $X$, $S$ and $Y$ form a Markov chain, in that order, we have
  \begin{align}
  I(X;Y)
  &= I(S;Y) - I(S;Y|X) + \underbrace{I(X;Y|S)}_{=0} \nonumber \\
  &= I(S;Y) - I(W;Y|X) \nonumber \\
  &= I(S;Y) - H(W|X) + H(W|X,Y) \nonumber \\
  &= I(S;Y) - H(W) + H(W|X,Y) \label{eq:proof-AMMC-mutual-1} \\
  &\leq C_{\text{\rm MMC}} - \log_q |\calT_{n \times m,t}| + H(W|X,Y) \label{eq:proof-AMMC-mutual-2}
  \end{align}
where (\ref{eq:proof-AMMC-mutual-1}) follows since $X$ and $W$ are independent.

  We now compute an upper bound on $H(W|X,Y)$. Let $R=\rank Y$ and write $Y = G\bar{Y}$, where $G \in \calT_{n \times R}$ and $\bar{Y} \in \calT_{R \times m}$. Note that
\begin{equation}\nonumber
  X + W = A^{-1} Y = A^{-1}G\bar{Y} = A^* \bar{Y}
\end{equation}
where $A^* = A^{-1}G$.
Since $\bar{Y}$ is full-rank, it must contain an invertible $R \times R$ submatrix. By reordering columns if necessary, assume that the left $R \times R$ submatrix of $\bar{Y}$ is invertible. Write $\bar{Y} = \mat{\bar{Y}_1 & \bar{Y}_2}$, $X = \mat{X_1 & X_2}$ and $W = \mat{W_1 & W_2}$, where $\bar{Y}_1$, $X_1$ and $W_1$ have $R$ columns, and $\bar{Y}_2$, $X_2$ and $W_2$ have $m-R$ columns. We have
\begin{equation}\nonumber
  A^* = (X_1 + W_1) \bar{Y}_1^{-1} \quad \text{and} \quad W_2 = A^* \bar{Y}_2 - X_2.
\end{equation}
It follows that $W_2$ can be computed if $W_1$ is known. Thus,
\begin{align}
H(W|X,Y)
&= H(W_1|X,Y) \leq H(W_1|R) \leq H(W_1|R=n) \nonumber \\
&\leq \log_q \sum_{i=0}^t |\calT_{n \times n,i}| \leq \log_q (t+1) |\calT_{n \times n,t}| \label{eq:proof-AMMC-bound-1}
\end{align}
where (\ref{eq:proof-AMMC-bound-1}) follows since $W_1$ may possibly be any $n \times n$ matrix with rank $\leq t$.

Applying this result in (\ref{eq:proof-AMMC-mutual-2}), and using (\ref{eq:bound-sum-gc}) and (\ref{eq:number-matrices-rank}), we have
\begin{align}
I(X,Y)
&\leq \log_q (n+1)\gc{m}{n} + \log_q (t+1)\frac{|\calT_{n\times t}|\gc{n}{t}}{|\calT_{n\times t}|\gc{m}{t}} \nonumber \\
&\leq \log_q (n+1)(t+1) \gc{m-t}{n-t} \label{eq:proof-AMMC-mutual-3} \\
  &\leq (m-n)(n-t) + \log_q 4(1+n)(1+t). \nonumber
\end{align}
where (\ref{eq:proof-AMMC-mutual-3}) follows from $\gc{m}{n} \gc{n}{t} = \gc{m}{t} \gc{m-t}{n-t}$, for $t \leq n \leq m$.
\end{proof}

We now develop a connection with the subspace approach of \cite{Kotter.Kschischang2008} that will be useful to obtain a lower bound on the capacity. From Section~\ref{sec:MMC}, we know that, in a multiplicative matrix channel, the receiver can only distinguish between transmitted subspaces.
Thus, we can equivalently express
\begin{equation}\nonumber
  C_{\text{\rm AMMC}} = \max_{p_X}\, I(\mathcal{X};\mathcal{Y})
\end{equation}
where $\mathcal{X}$ and $\mathcal{Y}$ denote the row spaces of $X$ and $Y$, respectively.

Using this interpretation, we can obtain the following lower bound on capacity.

\begin{theorem}\label{thm:AMMC-capacity-lower-bound}
  Assume $n \leq m$. For any $\epsilon \geq 0$, we have
  \begin{equation}\nonumber
    C_{\text{\rm AMMC}} \geq (m-n)(n-t-\epsilon t) - \log_q 4 - \frac{2tnm}{q^{1+\epsilon t}}.
  \end{equation}
\end{theorem}

In order to prove Theorem~\ref{thm:AMMC-capacity-lower-bound}, we need a few lemmas.

\begin{lemma}\label{lem:prob-low-rank}
  Let $X \in \Fq^{n \times m}$ be a matrix of rank $k$, and let $W \in \Fq^{n \times m}$ be a random matrix chosen uniformly among all matrices of rank $t$. If $k+t \leq \min\{n,m\}$, then
\begin{equation}\nonumber
  P[\rank(X+W) < k+t] < \frac{2t}{q^{\min\{n,m\}-k-t+1}}.
\end{equation}
\end{lemma}
\begin{proof}
  Write $X = X' X''$, where $X' \in \Fq^{n \times k}$ and $X'' \in \Fq^{k \times m}$ are full-rank matrices. We can generate $W$ as $W = W' W''$, where $W' \in \calT_{n \times t}$ and $W'' \in \calT_{t \times m}$ are chosen uniformly at random and independently from each other. Then we have
\begin{equation}\nonumber
  X + W = X'X''+W'W'' = \mat{X' & W'}\mat{X'' \\ W''}.
\end{equation}
Note that $\rank(X+W) = k+t$ if and only if the column spaces of $X'$ and $W'$ intersect trivially \emph{and} the row spaces of $X''$ and $W''$ intersect trivially. Let $P'$ and $P''$ denote the probabilities of these two events, respectively. By a simple counting argument, we have
\begin{align}
P'
&= \frac{(q^n-q^k)\cdots(q^n-q^{k+t-1})}{(q^n-1)\cdots(q^n-q^{t-1})} = \prod_{i=0}^{t-1}\frac{(1-q^{k-n+i})}{(1-q^{-n+i})} \nonumber \\
&> \prod_{i=0}^{t-1}(1-q^{k-n+i}) \geq (1-q^{k-n+t-1})^t \geq 1-tq^{k-n+t-1}. \nonumber
\end{align}
Similarly, we have $P'' > 1-tq^{k-m+t-1}$. Thus,
\begin{align}
P[\rank(X+W) < k+t] &< \frac{t}{q^{n-k-t+1}}+\frac{t}{q^{m-k-t+1}} \nonumber \\
&\leq \frac{2t}{q^{\min\{n,m\}-k-t+1}}. \nonumber
\end{align}
\end{proof}

For $\dim \mathcal{X} \leq n \leq m$, let $\mathcal{S}_{\mathcal{X},n}$ denote the set of all $n$-dimensional subspaces of $\Fq^m$ that contain a subspace $\mathcal{X} \subseteq \Fq^m$.

\begin{lemma}\label{lem:sphere-top-shell-size}
\begin{equation}\nonumber
  |\mathcal{S}_{\mathcal{X},n}| = \gc{m-k}{n-k}_q
\end{equation}
where $k = \dim \mathcal{X}$.
\end{lemma}
\begin{proof}
  By the fourth isomorphism theorem \cite{Dummit.Foote}, there is a bijection between $\mathcal{S}_{\mathcal{X},n}$ and the set of all $(n-k)$-dimensional subspaces of the quotient space $\Fq^m/\mathcal{X}$. Since $\dim \Fq^m/\mathcal{X} = m-k$, the result follows.
\end{proof}

We can now give a proof of Theorem~\ref{thm:AMMC-capacity-lower-bound}.

\begin{proof}[Proof of Theorem~\ref{thm:AMMC-capacity-lower-bound}]
  Assume that $X$ is selected from $\calT_{n \times m,k}$, where $k = n-(1+\epsilon)t$ and $\epsilon \geq 0$. Define a random variable $\AUXVAR$ as
\begin{equation}\nonumber
  \AUXVAR = \begin{cases}
    1 & \text{if } \dim \mathcal{Y} = \rank(X+W) = k+t \\
    0 & \text{otherwise.}
  \end{cases}
\end{equation}
Note that $\mathcal{X} \subseteq \mathcal{Y}$ when $\AUXVAR=1$.

By Lemma~\ref{lem:sphere-top-shell-size} and (\ref{eq:gauss-coeff-bound}), we have
\begin{equation}\nonumber
  H(\mathcal{Y} | \mathcal{X},\AUXVAR=1) \leq \log_q|\mathcal{S}_{\mathcal{X},n'}| \leq (m-n')t + \log_q 4
\end{equation}
where $n' = k+t$. Choosing $X$ uniformly from $\calT_{n \times m,k}$, we can also make $\mathcal{Y}$ uniform within a given dimension; in particular, \begin{equation}\nonumber
  H(\mathcal{Y}|\AUXVAR=1) = \log_q \gc{m}{n'}_q \geq (m-n')n'.
\end{equation}
It follows that
\begin{align}
I(\mathcal{X};\mathcal{Y}|\AUXVAR=1) &= H(\mathcal{Y}|\AUXVAR=1) - H(\mathcal{Y} | \mathcal{X},\AUXVAR=1) \nonumber \\
&\geq (m-n')(n'-t) - \log_q 4 \nonumber \\
&\geq (m-n)(n-t-\epsilon t) - \log_q 4. \nonumber
\end{align}

Now, using Lemma~\ref{lem:prob-low-rank}, we obtain
\begin{align}
I(\mathcal{X};\mathcal{Y})
&= I(\mathcal{X};\mathcal{Y},\AUXVAR) = I(\mathcal{X};\AUXVAR) + I(\mathcal{X};\mathcal{Y}|\AUXVAR) \nonumber \\
&\geq I(\mathcal{X};\mathcal{Y}|\AUXVAR) \nonumber \\
&\geq P[\AUXVAR=1]I(\mathcal{X};\mathcal{Y}|\AUXVAR=1) \nonumber \\
&\geq I(\mathcal{X};\mathcal{Y}|\AUXVAR=1) - P[\AUXVAR=0]nm \nonumber \\
&\geq (m-n)(n-t-\epsilon t) - \log_q 4 - \frac{2tnm}{q^{\epsilon t+1}}. \nonumber
\end{align}
\end{proof}

Note that, differently from the results of previous sections, Theorems~\ref{thm:AMMC-capacity-upper-bound} and \ref{thm:AMMC-capacity-lower-bound} provide only upper and lower bounds on the channel capacity. Nevertheless, it is still possible to compute exact expressions for the capacity of the AMMC in certain limiting cases.

\begin{corollary} \label{cor:AMMC-capacity-limit}
For $0 < \lambda = n/m \leq 1/2$ and $\tau = t/n$, we have
\begin{align}
\lim_{q \to \infty}\, C_{\text{\rm AMMC}} &= (m-n)(n-t) \label{eq:AMMC-capacity-limit-q} \\
\lim_{\substack{m \to \infty \\ n = \lambda m \\ t = \tau n}}\, \overline{C_{\text{\rm AMMC}}} &= (1 - \lambda)(1-\tau). \label{eq:AMMC-capacity-limit-m}
\end{align}
\end{corollary}
\begin{proof}
The fact that the values in (\ref{eq:AMMC-capacity-limit-q}) and (\ref{eq:AMMC-capacity-limit-m}) are upper bounds follows immediately from Theorem~\ref{thm:AMMC-capacity-upper-bound}. The fact that (\ref{eq:AMMC-capacity-limit-q}) is a lower bound follows immediately from Theorem~\ref{thm:AMMC-capacity-lower-bound} by setting $\epsilon = 0$. To obtain (\ref{eq:AMMC-capacity-limit-m}) from Theorem~\ref{thm:AMMC-capacity-lower-bound}, it suffices to choose $\epsilon$ such that $1/\epsilon$ grows sublinearly with $m$, e.g., $\epsilon = 1/\sqrt{m}$.
\end{proof}

Once again, note that (\ref{eq:AMMC-capacity-limit-q}) agrees with (\ref{eq:AMMC-capacity-limit-m}) if we consider the normalized capacity.

Differently from the MMC and the AMC, successful decoding in the AMMC does not (necessarily) allow recovery of all sources of channel uncertainty---in this case, the matrices $A$ and $W$. In general, for every observable $(X,Y)$ pair, there are many valid $A$ and $W$ such that $Y = A(X + W)$. Such coupling between $A$ and $W$ is reflected in extra term $H(W|X,Y)$ in (\ref{eq:proof-AMMC-mutual-1}), which provides an additional rate of roughly $(2n-t)t$ as compared to the straightforward lower bound $C_{\text{\rm AMMC}} \geq C_{\text{\rm MMC}} - \log_q |\calT_{n \times m,t}| \approx (m-n)n - (n+m-t)t$.

\begin{remark}
In \cite{Montanari.Urbanke2007}, the problem of communicating over the AMMC was addressed assuming a specific form of transmission matrices that contained an $n \times n$ identity header. Note that, if we consider such a header as being part of the channel (i.e., beyond the control of the code designer) then, with high probability, the resulting channel becomes equivalent to the AMC (see \cite{Montanari.Urbanke2007} for details). However, as a coding strategy for the AMMC, using such an $n \times n$ identity header results in a suboptimal input distribution, as the mutual information achieved is strictly smaller than the capacity. Indeed, the capacity-achieving distribution used in Theorem~\ref{thm:AMMC-capacity-lower-bound} and Corollary~\ref{cor:AMMC-capacity-limit} corresponds to transmission matrices of rank $n-(1+\epsilon)t$. This result shows that, for the AMMC, using headers is neither fundamental nor asymptotically optimal.
\end{remark}

\subsection{A Coding Scheme}

We now propose an efficient coding scheme that can asymptotically achieve (\ref{eq:AMMC-capacity-limit-q}) and (\ref{eq:AMMC-capacity-limit-m}). The scheme is based on a combination of channel sounding and error trapping strategies.

For a data matrix $\DATAMATRIX \in \Fq^{(n-v) \times (m-n)}$, where $v \geq t$, let the corresponding codeword be
\begin{equation}\nonumber
  X = \mat{0 \\ \bar{X}} = \mat{0_{v \times v} & 0_{v \times (n-v)} & 0_{v \times (m-n)} \\ 0_{(n-v) \times v} & I_{(n-v) \times (n-v)} & \DATAMATRIX}.
\end{equation}
Note that the all-zero matrices provide the error traps, while the identity matrix corresponds to the pilot symbols. Clearly, the rate of this scheme is $R = (n-v)(m-n)$.

Write the noise matrix $W$ as
\begin{equation}\nonumber
  W = BZ = \mat{B_1 \\ B_2} \mat{Z_1 & Z_2 & Z_3}
\end{equation}
where $B_1 \in \Fq^{v \times t}$, $B_2 \in \Fq^{(n-v) \times t}$, $Z_1 \in \Fq^{t \times v}$, $Z_2 \in \Fq^{t \times (n-v)}$ and $Z_3 \in \Fq^{t \times (m-n)}$.
The auxiliary matrix $S$ is then given by
\begin{equation}\nonumber
  S = X + W = \mat{B_1 Z_1 & B_1 Z_2 & B_1 Z_3 \\ B_2 Z_1 & I + B_2 Z_2 & \DATAMATRIX + B_2 Z_3}.
\end{equation}

Similarly as in Section~\ref{sec:AMC}, we define that the error trapping is successful if $\rank B_1 Z_1 = t$. Assume that this is the case. From Section~\ref{sec:AMC}, there exists some matrix $T \in \calT_{n \times n}$ such that
\begin{equation}\nonumber
  TS = \mat{B_1 Z_1 & B_1 Z_2 & B_1 Z_3 \\ 0 & I & \DATAMATRIX} = \mat{B_1 & 0 \\ 0 & I} \mat{Z_1 & Z_2 & Z_3 \\ 0 & I & \DATAMATRIX}.
\end{equation}
Note further that
\begin{equation}\nonumber
  \RRE{\mat{Z_1 & Z_2 & Z_3 \\ 0 & I & \DATAMATRIX}} = \mat{\tilde{Z}_1 & 0 & \tilde{Z}_3 \\ 0 & I & \DATAMATRIX}
\end{equation}
for some $\tilde{Z}_1 \in \Fq^{t \times v}$ in RRE form and some $\tilde{Z}_3 \in \Fq^{t \times (m-n)}$.
It follows that
\begin{align}
\RRE{S}
&= \RRE{\mat{B_1 & 0 \\ 0 & I} \mat{Z_1 & Z_2 & Z_3 \\ 0 & I & \DATAMATRIX}} \nonumber \\
&= \mat{\tilde{Z}_1 & 0 & \tilde{Z}_3 \\ 0 & I & \DATAMATRIX \\ 0 & 0 & 0} \nonumber
\end{align}
where the bottom $v-t$ rows are all-zeros.

Since $A$ is invertible, we have $\RRE{Y} = \RRE{S}$, from which $\DATAMATRIX$ can be readily obtained. Thus, decoding amounts to performing Gauss-Jordan elimination on $Y$. It follows that the complexity of the scheme is $O(n^2m)$ operations in $\Fq$.

The probability that the error trapping is not successful, i.e., $\rank B_1 Z_1 < t$, was computed in Section~\ref{sec:AMC}. Let $\hat{A}$ correspond to the first $n$ columns of $Y$. Note that $\rank B_1 Z_1 = t$ if and only if $\rank \hat{A} = n-v+t$. Thus, when the error trapping is not successful, the receiver can easily detect this event by looking at $\RRE{Y}$ and then declare a decoding failure. It follows that the scheme has zero error probability and probability of failure given by (\ref{eq:AMC-failure-prob}).

\begin{theorem}\label{thm:AMMC-coding-achievability}
  The proposed coding scheme can asymptotically achieve (\ref{eq:AMMC-capacity-limit-q}) and (\ref{eq:AMMC-capacity-limit-m}).
\end{theorem}
\begin{proof}
Using (\ref{eq:AMC-failure-prob}) and the same argument as in the proof of Proposition~\ref{prop:AMC-coding-achievability}, we can set a suitable $v$ in order to achieve arbitrarily low gap to capacity while maintaining an arbitrary low probability of failure, for both cases where $q \to \infty$ or $m \to \infty$.
\end{proof}

\section{Extensions}
\label{sec:extensions}

In this section, we discuss possible extensions of the results and models presented in the previous sections.

\subsection{Dependent Transfer Matrices}

As discussed in Section~\ref{sec:AMMC}, the AMMC is equivalent to a channel of the form (\ref{eq:prob-basic-channel-law}) where $A \in \calT_{n \times n}$ and $D \in \calT_{n \times t}$ are chosen uniformly at random and independently from each other. Suppose now that the channel is the same, except for the fact that $A$ and $D$ are not independent. It should be clear that the capacity of the channel cannot be smaller than that of the AMMC. For instance, one can always convert this channel into an AMMC by employing randomization at the source. (This is, in fact, a natural procedure in any random network coding system.) Let $X = TX'$, where $T \in \calT_{n \times n}$ is chosen uniformly at random and independent from any other variables. Then $A' = A T$ is uniform on $\calT_{n \times n}$ and independent from $D$. Thus, the channel given by $Y = A'X' + DZ$ is an AMMC.

Note that our coding scheme does not rely on any particular statistics of $A$ given $X$ and $W$ (except the assumption that $A$ is invertible) and therefore works unchanged in this case.

\subsection{Transfer Matrix Invertible but Nonuniform}

The model for the AMMC assumes that the transfer matrix $A \in \calT_{n \times n}$ is chosen uniformly at random. In a realistic network coding system, the transfer matrix may be a function of both the network code and the network topology, and therefore may not have a uniform distribution. Consider the case where $A$ is chosen according to an arbitrary probability distribution on $\calT_{n \times n}$. It should be clear that the capacity can only increase as compared with the AMMC, since less ``randomness'' is introduced in the channel. The best possible situation is to have a constant $A$, in which case the channel becomes exactly an AMC.

Again, note that our coding scheme for the AMMC is still applicable in this case.


\subsection{Nonuniform Packet Errors}
\label{ssec:nonuniform-packet-errors}

When expressed in the form (\ref{eq:prob-basic-channel-law}), the models for both the AMC and the AMMC assume that the matrix $Z$ is uniformly distributed on $\calT_{n \times t}$. In particular, each error packet is uniformly distributed on $\Fq^{1 \times m} \setminus \{0\}$. In a realistic situation, however, it may be the case that error packets of low weight are more likely to occur. Consider a model identical to the AMC or the AMMC except for the fact that the matrix $Z$ is chosen according to an arbitrary probability distribution on $\calT_{t \times m}$. Once again, it should be clear that the capacity can only increase. Note that the exact capacity in Proposition~\ref{prop:AMC-capacity} and the upper bound of Theorem~\ref{thm:AMMC-capacity-upper-bound} can be easily modified to account for this case (by replacing $\log_q |\calT_{n \times m, t}|$ with the entropy of $W$).

Although our coding scheme in principle does not hold in this more general case, we can easily convert the channel into an AMC or AMMC by applying a random transformation at the source (and its inverse at the destination). Let $X = X' T$, where $T \in \calT_{m \times m}$ is chosen uniformly at random and independent from any other variables. Then
\begin{equation}\nonumber
  Y' = Y T^{-1} = (AX + DZ)T^{-1} = AX' + DZ'
\end{equation}
where $Z' = Z T^{-1}$. Since $\calT_{m \times m}$ acts (by right multiplication) transitively on $\calT_{t \times m}$, we have that $Z'$ is uniform on $\calT_{t \times m}$. Thus, we obtain precisely an AMMC (or AMC) and the assumptions of our coding scheme hold.

Note, however, that, depending on the error model, the capacity may be much larger than what can be achieved by the scheme described above. For instance, if the rows of $Z$ are constrained to have weight at most $s$ (otherwise chosen, say, uniformly at random), then the capacity would increase by approximately $\left(m - s - \log_q \binom{m}{s}\right)t$, which might be a substantial amount if $s$ is small.

\subsection{Error Matrix with Variable Rank ($\leq t$)}

The model we considered for the AMC and the AMMC assumes an error matrix $W$ whose rank is known and equal to $t$. It is useful to consider the case where $\rank W$ is allowed to vary, while still bounded by $t$. More precisely, we assume that $W$ is chosen uniformly at random from $\calT_{n \times m,R}$, where $R \in \{0,\ldots,t\}$ is a random variable with probability distribution $P[R = r] = p_r$.

Since
\begin{align}
H(W)
&= H(W,R) = H(R) + H(W|R) \nonumber \\
&= H(R) + \sum_{r} p_r H(W|R=r) \nonumber \\
&= H(R) + \sum_{r} p_r \log_q|\calT_{n \times m,r}| \nonumber \\
&\leq H(R) + \log_q|\calT_{n \times m,t}|, \nonumber
\end{align}
we conclude that the capacities of the AMC and the AMMC may be reduced by at most $H(R) \leq \log_q(t+1)$. This loss is asymptotically negligible for large $q$ and/or large $m$, so the expressions (\ref{eq:AMC-capacity-limit-q}), (\ref{eq:AMC-capacity-limit-m}), (\ref{eq:AMMC-capacity-limit-q}) and (\ref{eq:AMMC-capacity-limit-m}) remain unchanged.

The steps for decoding and computing the probability of error trapping failure also remain the same, provided we replace $t$ by $R$. The only difference is that now decoding errors may occur. More precisely, suppose that $\rank B_1 Z_1 = t' < t$. A necessary condition for success is that $\rank B_1 Z = \rank B Z_1 = t'$. If this condition is not satisfied, then a decoding failure is declared. However, if the condition is true, then the decoder cannot determine whether $t' = R < t$ (an error trapping success) or $t' < R \leq t$ (an error trapping failure), and must proceed assuming the former case. If the latter case turns out to be true, we would have an undetected error. Thus, for this model, the expression (\ref{eq:AMC-failure-prob}) gives a bound on the probability that decoding is not successful, i.e., that either an error or a failure occurs.

\subsection{Infinite-Packet-Length Channel and Infinite-Batch-Size Channel}
\label{ssec:other-infinite-channels}

We now extend our results to the infinite-packet-length AMC and AMMC and the infinite-batch-size AMC. (Note that, as pointed out in Section~\ref{sec:MMC}, there is little justification to consider an infinite-batch-size AMMC.) From the proof of Propositon~\ref{prop:AMC-capacity} and the proof of Corollary~\ref{cor:AMMC-capacity-limit}, it is straightforward to see that
\begin{align}
 \lim_{m \to \infty}\, \overline{C_{\text{\rm AMMC}}} = \lim_{m \to \infty}\, \overline{C_{\text{\rm AMC}}}  &= (n-t)/n  \nonumber \\
 \lim_{n \to \infty}\, \overline{C_{\text{\rm AMC}}} &= (m-t)/m. \nonumber
\end{align}

It is \emph{not} straightforward, however, to obtain capacity-achieving schemes for these channels. The schemes described in Sections~\ref{sec:AMC} and~\ref{sec:AMMC} for the infinite-rank AMC and AMMC, respectively, use an error trap whose size (in terms of columns \emph{and} rows) grows proportionally with $m$ (or $n$). While this is necessary for achieving vanishingly small error probability, it also implies that these schemes are not suitable for the infinite-packet-length channel (where $m \to \infty$ but not $n$) or the infinite-batch-size channel (where $n \to \infty$ but not $m$).

In these situations, the proposed schemes can be adapted by replacing the data matrix and part of the error trap with a \emph{maximum-rank-distance} (MRD) code \cite{Gabidulin1985}. Consider first an infinite-packet-length AMC. Let the transmitted matrix be given by
\begin{equation}\label{eq:AMC-MRD-error-trap}
 X = \mat{0_{n \times v} & x}
\end{equation}
where $x \in \Fq^{n \times (m-v)}$ is a codeword of a matrix code $\calC$. If (column) error trapping is successful then, under the terminology of \cite{Silva++2008}, the decoding problem for $\calC$ amounts to the correction of $t$ \emph{erasures}. It is known that, for $m-v \geq n$, an MRD code $\calC \subseteq \Fq^{n \times (m-v)}$ with rate $(n-t)/n$ can correct exactly $t$ erasures (with zero probability of error) \cite{Silva++2008}. Thus, decoding fails if and only if column trapping fails.

Similarly, for an infinite-batch-size AMC, let the transmitted matrix be given by
\begin{equation}\nonumber
 X = \mat{0_{v \times m} \\ x}
\end{equation}
where $x \in \Fq^{(n-v) \times m}$ is a codeword of a matrix code $\calC$. If (row) error trapping is successful then, under the terminology of \cite{Silva++2008}, the decoding problem for $\calC$ amounts to the correction of $t$ \emph{deviations}. It is known that, for $n-v \geq m$, an MRD code $\calC \subseteq \Fq^{(n-v) \times m}$ with rate $(m-t)/m$ can correct exactly $t$ deviations (with zero probability of error) \cite{Silva++2008}. Thus, decoding fails if and only if row trapping fails.

Finally, for the infinite-packet-length AMMC, it is sufficient to prepend to (\ref{eq:AMC-MRD-error-trap}) an identity matrix, i.e.,
\begin{equation}\nonumber
 X = \mat{I_{n \times n} & 0_{n \times v} & x}.
\end{equation}
The same reasoning as for the infinite-packet-length AMC applies here, and the decoder in \cite{Silva++2008} is also applicable in this case.

For more details on the decoding of an MRD code combined with an error trap, we refer the reader to \cite{Silva.Kschischang2009:KeyBased-CWIT}. The decoding complexity is in $O(tn^2 m)$ and $O(t m^2 n)$ (whichever is smaller) \cite{Silva++2008}.

In all cases, the schemes have probability of error upper bounded by $t/q^{1 + v-t}$ and therefore are capacity-achieving.

\section{Conclusions}
\label{sec:conclusion}

We have considered the problem of reliable communication over certain additive matrix channels inspired by network coding. These channels provide a reasonable model for both coherent and random network coding systems subject to random packet errors. In particular, for an additive-multiplicative matrix channel, we have obtained upper and lower bounds on capacity for any channel parameters and asymptotic capacity expressions in the limit of large field size and/or large matrix size; roughly speaking, we need to use $t$ redundant packets in order to be able to correct up to $t$ injected error packets. We have also presented a simple coding scheme that achieves capacity in these limiting cases while requiring a significantly low decoding complexity; in fact, decoding amounts simply to performing Gauss-Jordan elimination, which is already the standard decoding procedure for random network coding. Compared to previous work on correction of adversarial errors (where approximately $2t$ redundant packets are required), the results of this paper show an improvement of $t$ redundant packets that can be used to transport data, if errors occur according to a probabilistic model.

Several questions remain open and may serve as an interesting avenue for future research:
\begin{itemize}
  \item Our results for the AMMC assume that the transfer matrix $A$ is always nonsingular. It may be useful to consider a model where $\rank A$ is a random variable. Note that, in this case, one cannot expect to achieve reliable (and efficient) communication with a one-shot code, as the channel realization would be unknown at the transmitter. Thus, in order to achieve capacity under such a model (even with arbitrarily large $q$ or $m$), it is strictly necessary to consider multi-shot codes.

  \item As pointed out in Section~\ref{ssec:nonuniform-packet-errors}, our proposed coding scheme may not be even close to optimal when packet errors occur according to a nonuniform probability model. Especially in the case of low-weight errors, it is an important question how to approach capacity with a low-complexity coding scheme. It might also be interesting to know whether one-shot codes are still useful in this case.

  \item Another important assumption of this paper is the bounded number of $t<n$ packet errors. What if $t$ is unbounded (although with a low number of errors being more likely than a high number)? While the capacity of such a channel may not be too hard to approximate (given the results of this paper), finding a low-complexity coding scheme seems a very challenging problem.
\end{itemize}

\section*{Acknowledgements}
  We would like to thank the associate editor and the anonymous reviewers for their helpful comments.

\begin{IEEEbiographynophoto}{Danilo Silva}
(S'06) received the B.Sc. degree from the Federal University of Pernambuco, Recife, Brazil, in 2002, the M.Sc. degree from the Pontifical Catholic University of Rio de Janeiro (PUC-Rio), Rio de Janeiro, Brazil, in 2005, and the Ph.D. degree from the University of Toronto, Toronto, Canada, in 2009, all in electrical engineering.

He is currently a Postdoctoral Fellow at the University of Toronto. His research interests include channel coding, information theory, and network coding.
\end{IEEEbiographynophoto}

\begin{IEEEbiographynophoto}{Frank R. Kschischang}
(S'83--M'91--SM'00--F'06) received the B.A.Sc. degree (with honors) from the University of British Columbia, Vancouver, BC, Canada, in 1985 and the M.A.Sc. and Ph.D. degrees from the University of Toronto, Toronto, ON, Canada, in 1988 and 1991, respectively, all in electrical engineering.

He is a Professor of Electrical and Computer Engineering and Canada Research Chair in Communication Algorithms at the University of Toronto, where he has been a faculty member since 1991. During 1997--1998, he was a Visiting Scientist at the Massachusetts Institute of Technology, Cambridge, and in 2005 he was a Visiting Professor at the ETH, Z{\"u}rich, Switzerland. His research interests are focused on the area of channel coding techniques.

Prof. Kschischang was the recipient of the Ontario Premier's Research Excellence Award. From 1997 to 2000, he served as an Associate Editor for Coding Theory for the \textsc{IEEE Transactions on Information Theory}. He also served as Technical Program Co-Chair for the 2004 IEEE International Symposium on Information Theory (ISIT), Chicago, IL, and as General Co-Chair for ISIT 2008, Toronto.
\end{IEEEbiographynophoto}

\begin{IEEEbiographynophoto}{Ralf K{\"o}tter}
(S'92-M'95-SM'06-F'09) received a Diploma in Electrical Engineering from the Technical University Darmstadt, Germany in 1990 and a Ph.D. degree from the Department of Electrical Engineering at Link{\"o}ping University, Sweden.

From 1996 to 1997, he was a Visiting Scientist at the IBM Almaden Research Laboratory in San Jose, CA. He was a Visiting Assistant Professor at the University of Illinois at Urbana-Champaign and a Visiting Scientist at CNRS in Sophia-Antipolis, France, from 1997 to 1998. In the years 1999-2006 he was member of the faculty of the University of Illinois at Urbana-Champaign, where his research interests included coding and information theory and their application to communication systems. In 2006 he joined the faculty of the Technische Universit{\"a}t M{\"u}nchen, Munich, Germany, as the Head of the Institute for Communications Engineering.

He served as an Associate Editor for both the \textsc{IEEE Transactions on Communications} and the \textsc{IEEE Transactions on Information Theory}.
He received an IBM Invention Achievement Award in 1997, an NSF CAREER Award in 2000, an IBM Partnership Award in 2001, and a 2006 XEROX award for faculty research. He is co-recipient of the 2004 Information Theory Society Best Paper Award, of the 2004 IEEE Signal Processing Magazine Best Paper Award, and of the 2009 Joint Communications Society and Information Theory Society Best Paper Award. He received the Vodafone Innovationspreis in 2008.

Ralf K\"otter died in February, 2009.
\end{IEEEbiographynophoto}

\end{document}